\documentclass{sig-alternate}
\usepackage{graphicx}
\usepackage{url}
\usepackage[expansion=true,kerning,spacing,protrusion=true]{microtype}

\begin{document}
%
\conferenceinfo{MSR}{'11 Honolulu, Hawaii USA}

\title{Temporal Analysis of Literary and Programming Prose}

\numberofauthors{3} 

\author{
\alignauthor
Brian Michalski\\
       \affaddr{Rensselaer Polytechnic Institute}\\
       \affaddr{110 8th Street}\\
       \affaddr{Troy, NY}\\
       \email{bmichalski@gmail.com}
\alignauthor
Mukkai Krishnamoorthy\\
       \affaddr{Rensselaer Polytechnic Institute}\\
       \affaddr{110 8th Street}\\
       \affaddr{Troy, NY}\\
       \email{mskmoorthy@gmail.com}
\alignauthor
Tsz-Yam Lau\\
       \affaddr{Rensselaer Polytechnic Institute}\\
       \affaddr{110 8th Street}\\
       \affaddr{Troy, NY}\\
       \email{laut@cs.rpi.edu}
}
\maketitle
\begin{abstract}
Literary works reference a variety of globally shared themes including well-known people, events, and time periods. It is particularly interesting to locate patterns that are either invariant across time or exhibit a characteristic change across time, as they could imply something important about society that those works record. This paper suggests the use of Google n-gram viewer as a fast prototyping method for examining time-based properties over a rich sample of literary prose. Using this method, we find that some repeating periods of time, like Sunday, are referenced disproportionally, allowing us to pose questions such as why a day like Thursday is so unpopular. Furthermore, by treating software as a work of prose, we can apply a similar analysis to open-source software repositories and explore time-based relations in commit logs. Doing a simple statistical analysis on a few temporal keywords in the log records, we reinforce and weaken a few beliefs on how college students approach open source software. Finally, we help readers working on their own temporal analysis by comparing the fundamental differences between literary works and code repositories, and suggest blog or wiki as recently-emerging works.
\end{abstract}




\newpage

\section{Introduction}
In the past, discovery of issues recorded in written works depended on lots of manpower reading through printed or even hand-written scripts. Nowadays, the task is made easier with digitization: it has become easier to look for particular keywords from e-books circulated in digital format. It is also much easier to work with old literary works after digitizing them with the scanning and optical character recognition (OCR) technology.

This paper brings our readers' attention to temporal analysis of literary works, in which we look for characteristic patterns across time. It could be an invariance which stands the test of time. It could be a trend which is changing. Below are some examples:
\begin{itemize}
\item Appearance of temporal indicators (such as the days of the week) across prose: In novels, it could mean the time of a setting. In news reports, their appearance indicates the time instances at which interesting events occurred.
\item Evolution across the same prose - when the book/article is published, any invariance/change in the word used.
\end{itemize}

In an age in which digital content is evolving at an increasingly rapid pace, we look back at the seemingly analog form of literary repositories and look to extrapolate ideas that may also hold true for extremely technical prose, that of software repositories. Software source code written in a programming language such as Ruby, C, or Python code may not read like the work of Shakespeare to an average reader; however, the additional data stored in repository history logs provides a level of evolutionary detail that we are not fortunate enough to access in the traditional literary world.

Literature and software code share many of the same characteristics and lend them well to being compared.  Both works, in a general sense, need to be compiled or interpreted on a per-processor basis to be used in their intended form: the two forms follow well developed patterns; in software those patterns are validated by a compiler and in literature they may be validated by a dictionary or grammar guide.  By treating literature and software as fundamentally the same type of repository, we conduct several experiments and infer the properties that may be shared between the literary and software sources.

We start out by presenting an analysis of both literary and programming prose. The approach is mainly hypothesis testing, which depends on researchers initiating certain proposals. We then attempt to use statistics to see if they are statistically significant or not. It is unlike data mining, in which we mainly ask the computer program to cluster the items to discover possible patterns itself. This simpler practice allows us to obtain statistics using existing tools, so as to inspire those without programming knowledge to initiate their own creative designs. We also present a few interesting results that are worth further thinking. This paper is organized as follows. In Section 2, we detail the methodology. In Section 3, we present the interesting findings that we have found from literary and programming prose, before we conclude and discuss possible future works in Section 4.

\section{Methodology}
\subsection{Literary Works}
For literary works, we chose measurable components such as numerical references to years and English week day names. By choosing generally universal search terms, we focus our efforts away from the parsing and extraction to the inferences and comparisons between samples of repositories and that of the whole. We can fit those queries in the web interface described below.

To look for recurring keywords in literary writings, we make use of the n-gram viewer \cite{Google2010} in Google Books. The company has precompiled the set of one- to five- grams from over a million of books that they scanned and were published between the years 1800 and 2000. It keeps a representative sample of literary works and provides a precompiled dataset for fast queries. It also offers a web page \footnote{\texttt{http://books.google.com/ngrams}} where amateurs can plot the relative frequencies of one or more n-grams over time. This feature has already been used to analyze cultural trends analysis \cite{Michel10}.

\begin{table}[tbp]
\centering
\caption{Books with multiple editions in a sample of 1000 books}
\begin{tabular}{|c|c|} \hline
Number of Editions&Books\\ \hline
2 & 29\\ \hline
3 & 6\\ \hline
4 & 3\\ \hline
5 & 1\\ \hline
10 & 1\\ \hline
\end{tabular}
\label{editions_table}
\end{table}

Evolution across the same literary prose, however, is not interesting in general. Literature, at least the publicly available sources these authors are privy to, rarely store transactional data or any temporal data before release.  The development of literary works, and books in particular, is much harder to track over time as a result.  We are left to look only at what publishers and authors formally print and release as a new edition or re-release of a work.  In a sample of 1000 public domain volumes supplied by Google \cite{booksearchdata}, we found 40 sets of books with more than one revision, identified based on the similarity of their titles and authors. Only 11 out of those 40 books had more than two revisions, as shown in Table \ref{editions_table}.  The small sample size of books with more than two revisions (only slightly larger than 1\% of our sample) limits the number of meaningful analyses we can perform using literary revisions.  A larger sample of works, particularly those authored in more recent years, where it is believed revisioning is more likely to occur due to the digital nature of the source material, may provide more fruitful results.

\subsection{Software}
Keyword search in source code can be done through specialized search engines \footnote{For example, \texttt{http://www.google.com/codesearch}} where one can search for code snippets. In a sense, code search is more similar to book search or literature search, as search is carried out based more on key words than on links. 



However, unlike books which often mention temporal indicators such as days of the week as events in both fiction and non-fiction,   temporal keywords are seldom mentioned in the programs (as variable names or functions) unless they are doing calendrical computations. 

In contrast, we can often extract much richer development history of code than literacy counterparts. A version control system \cite{Wikipedia_rcs} is often used in large code development projects to manage changes to code stored in digital format. It associates each change with a specific date and time. For program code, such change logs are named commits. A commit refers to the idea of making an often subject-based grouping of tentative changes, in this context changes of code in a developer's local computer, permanent by uploading the changes to a code repository \cite{Wikipedia_commit}. This is the final step that a programmer does when he is satisfied with the modifications he has made on the local code. As a result, a commit can be treated as a record of progress.

One may access the commit logs to explore this temporal feature of code at the commit level. For example, with \textit{git} version control system, he can use the \texttt{git log} functionality \footnote{\texttt{http://learn.github.com/p/log.html}}. With a simple command, he can obtain the date, time and author of each commit.

\section{Findings}
\subsection{Literary Works}
To test the consistency property of certain temporal keywords over an extended period of time given a reasonably static social factor, we explore the specific references to those keywords in a data set that spans hundreds of years. We accomplish this through use of the Google Books 1-grams English data set mentioned in the previous section. Table \ref{bin_size_table} suggests the distribution of the publication years of the books being used to compile the statistics.

\begin{table}[tbp]
\centering
\caption{Distribution of publication years of the books}
\begin{tabular}{|c|c|} \hline
\textbf{Years} & \textbf{Number of books} \\ \hline
1520 - 1699    & 1,243\\ \hline
1700 - 1799    & 44,059\\ \hline
1800 - 1899    & 5,518,213\\ \hline
1900 - 2008    & 31,823,074\\ \hline
\end{tabular}
\label{bin_size_table}
\end{table}

\subsubsection{Days of the week}
\begin{figure*}[htb]
\centering
\includegraphics[width=0.9\textwidth]{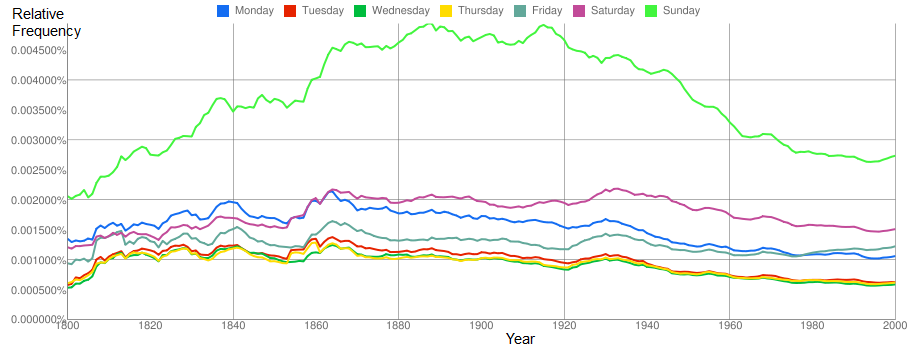}
\caption{Relative frequencies of day of the week mentions across time}
\label{day_of_week_label}
\end{figure*}

This involves accessing the number of references by case insensitive listings for the singular form of the seven days of the week:
\textit{Sunday}, \textit{Monday}, \textit{Tuesday}, \textit{Wednesday}, \textit{Thursday}, \textit{Friday}, and \textit{Saturday}. The resulting relative frequency plot is shown in Figure \ref{day_of_week_label}.



Leaving aside speculations regarding the literary significance of Sundays at nearly double frequency or the lack of popularity Tuesday through Thursday hold in our sample to our literary counterparts, our data indicates much broader trends that hold true for a significant period of time.  The consistent result indicates the effect of sampling by century was minimal, despite the significant differences caused by spelling or optical character recognition challenges, notably the medial s that declined significantly in the 1800s. 

\subsubsection{Months of the year}
\begin{figure*}[htb]
\centering
\includegraphics[width=0.9\textwidth]{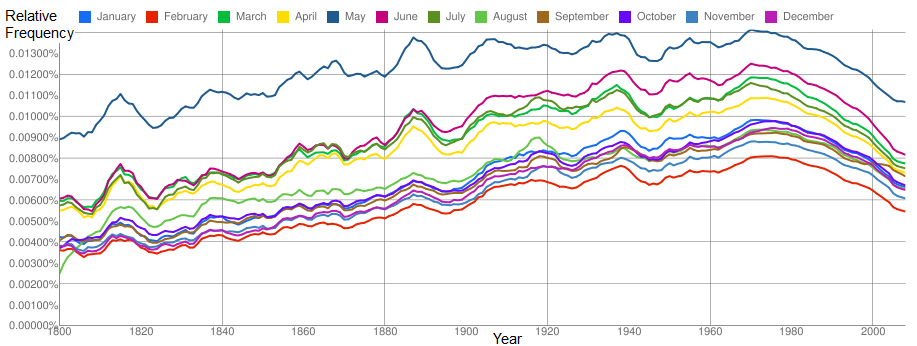}
\caption{Relative frequencies of month of the year mentions across time}
\label{month_of_year_label}
\end{figure*}

We find their relative frequencies remain consistent, as shown in Figure \ref{month_of_year_label}, across centuries and in the overall distribution.   While our repository of English literature spans over 450 years, this specific trend and presumably many others remains remarkably consistent given the wide variety and literary development during that time period.

Relative frequencies of the months show similar consistency over time too. Check Figure \ref{month_of_year_label}. Excluding May and June which could also be used as names and verbal auxiliary and thus expected to show up more often, we observe a few months that are consistently occurring more than other months, such as July.

\subsubsection{Relative temporal indictators}
\begin{figure*}[htb]
\centering
\includegraphics[width=0.9\textwidth]{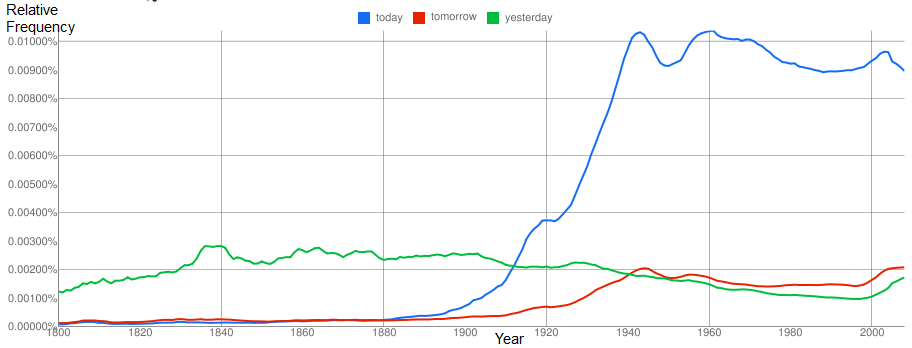}
\caption{Relative frequencies of ``today'', ``yesterday'' and ``tomorrow'' across time}
\label{relative_label}
\end{figure*}

Finally, we look at relative temporal indicators including ``today'', ``yesterday'' and ``tomorrow''. See Figure \ref{relative_label}. Before around 1800, ``yesterday'' dominated. Afterwards, the usage of ``today'' boomed and outnumbered the others. We also observe a similar rise of ``tomorrow''. Though the premium is not as large as that of ``today'', it also consistently outnumbers ``yesterday''.

\subsection{Software}
In this section, we analyze the git repositories maintained by the Rensselaer Center of Open Source software (RCOS) in Rensselaer Polytechnic Institute (RPI) \footnote{\texttt{http://rcos.rpi.edu/}}. RCOS is a group of RPI students who work on a variety of self-initiated open source projects. They participate for course credits or a stipend. We pick a set of eight repositories as shown in Table \ref{projectdetails}. The chosen projects are under active developments as indicated by their large numbers of commits.

\begin{table}[tbp]
\centering
\caption{The eight open source projects under analysis}
\small{
\begin{tabular}{|c|c|c|c|c|} \hline
\textbf{Project name} & \textbf{From} & \textbf{To} & \textbf{Days} & \textbf{Commits} \\ \hline
Briefcase             & 09/17/11      & 12/07/11    & 81            & 132     \\ \hline
Convalot              & 09/12/10      & 06/25/11    & 286           & 406     \\ \hline
Milkyway              & 07/31/08      & 12/14/11    & 1231          & 2436    \\ \hline
MobileNotifier        & 09/11/10      & 10/17/11    & 401           & 450     \\ \hline
Notebook              & 11/18/10      & 04/06/11    & 139           & 326     \\ \hline
RPIDirectory          & 08/31/11      & 09/30/11    & 30            & 193     \\ \hline
ShuttleTracking       & 08/02/09      & 11/04/11    & 824           & 207     \\ \hline
YACS                  & 09/14/11      & 12/14/11    & 91            & 273     \\ \hline
\end{tabular}
}
\label{projectdetails}
\end{table}

\subsubsection{Hours of the day}
\begin{figure*}[htb]
\centering
\includegraphics[width=0.65\textwidth]{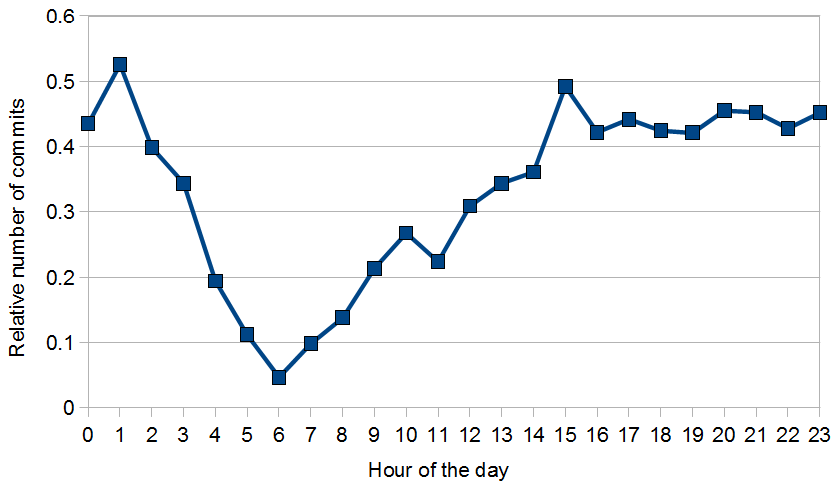}
\caption{Relative frequencies of commits at different time of a day}
\label{software_hours}
\end{figure*}

We start by looking at the distribution of the commits over the 24 hours of the day. We are interested in the time of the day when the students work, so we take the programmers' commit time without converting them to time corresponding to the time zone of the school. Instead of summing up the actual number of commits, we add up the proportion of commits of all the projects in the same hour together and compile Figure \ref{software_hours}. This prevents projects with huge total number of commits from dominating the statistics and thus equalizes the influence of individual projects. 

As shown in Figure \ref{software_hours}, the proportion of commits remains at a relatively high level from the afternoon till early morning the next day. A trough is observed at around 6 a.m. This matches our expectation that students tend to stay late at night or even till early in the morning to code. This also confirms the importance of shifting actual time to honor the concept of a day of a typical person. Students usually define that a day ends when they go to bed, and it can be extended beyond chronological midnight.

\subsubsection{Days of the week}
After adjusting for the extended night hours, we can then analyze the commit data by the unit of days. The students are asked to report their progress every Friday at 4 P.M. during regular semester and at noon during summer. A question is to see if students' work (and hence commits) are concentrated on the few days before the project progress presentations. Statistically, we need to check if the number of commits is significantly larger on Wednesdays and Thursdays.

To do this, we compile the proportion of commits at different days of the week, as shown in Table \ref{project-commit-day-of-the-week}.

\begin{table*}[tbp]
\centering
\caption{Proportion of commits in different days of the week (a day starts at 6.00am and ends at 5.59am the next day)}
\begin{tabular}{|c|c|c|c|c|c|c|c|} \hline
\textbf{Project name} & \textbf{Mon} & \textbf{Tue} & \textbf{Wed} & \textbf{Thurs} & \textbf{Fri} & \textbf{Sat} & \textbf{Sun} \\ \hline
Briefcase             & 19.70\%      & 13.64\%      & 11.36\%      & 15.15\%        &  6.82\%      & 21.97\%      & 11.36\% \\ \hline
Convalot              & 13.79\%      & 10.34\%      & 14.29\%      & 12.07\%        & 20.20\%      & 20.69\%      &  8.62\% \\ \hline
Milkyway              & 14.70\%      & 16.17\%      & 15.60\%      & 15.15\%        & 14.57\%      & 12.97\%      & 10.84\% \\ \hline
MobileNotifier        & 12.44\%      & 22.44\%      &  8.44\%      & 14.67\%        & 11.11\%      & 17.78\%      & 13.11\% \\ \hline
Notebook              & 10.74\%      & 10.43\%      & 19.63\%      & 16.26\%        & 15.34\%      & 14.42\%      & 13.19\% \\ \hline
RPIDirectory          &  3.63\%      & 31.61\%      & 14.51\%      & 25.39\%        & 17.62\%      &  1.04\%      &  6.22\% \\ \hline
ShuttleTracking       &  8.21\%      & 13.04\%      & 31.40\%      &  6.76\%        & 16.43\%      &  8.21\%      & 15.94\% \\ \hline
YACS                  & 15.02\%      & 12.82\%      & 11.72\%      & 19.05\%        & 16.48\%      &  7.69\%      & 17.22\% \\ \hline
\hline
Mean deviation from 14.29\% & -2.01\% & 2.03\%  &  1.58\% & 1.28\%  & 0.53\%  & -1.19\% & -2.22\% \\ \hline
Standard deviation          &  4.85\% & 7.29\%  &  7.10\% & 5.34\%  & 4.14\%  & 7.16\%  & 3.62\%  \\ \hline
Standard error              &  1.71\% & 2.58\%  &  2.51\% & 1.89\%  & 1.47\%  & 2.53\%  & 1.28\%  \\ \hline
95\% CI max                 &  1.35\% & 7.08\%  &  6.50\% & 4.98\%  & 3.41\%  & 3.77\%  & 0.29\%  \\ \hline
95\% CI min                 & -5.37\% &-3.02\%  & -3.33\% &-2.43\%  &-2.34\%  &-6.15\%  &-4.14\% \\ \hline 
\end{tabular}
\label{project-commit-day-of-the-week}
\end{table*}


Then we compute how these figures are deviated from the expected value 1/7 = 14.29\%, which is the expected proportion if the commits are distributed evenly across the seven days of the week. All the 95\% confidence intervals (C.I.) include 0, which means we are not confident to say any deviation is significant relative to the expected evenly-distributed case.   

\subsubsection{Coding session length} 
When interpreting commit, note that a commit just means a confirmation. The hours of work behind it are impossible to tell. However, it makes sense to cluster consecutive commits which are temporarily close with their immediate neighbor and estimate approximate session length. 
Table \ref{projectdurations} shows the average length of the sessions. The grand mean is around 4.5 hours. 

\begin{table}[tbp]
\centering
\caption{Average session durations of the eight projects under investigation}
\begin{tabular}{|c|c|c|c|c|} \hline
\textbf{Project name} & \textbf{Duration} \\ \hline
Briefcase             & 3 hr 5 min    \\ \hline
Convalot              & 4 hr 1 min    \\ \hline
Milkyway              & 4 hr 55 min \\ \hline
MobileNotifier        & 4 hr 2 min  \\ \hline
Notebook              & 6 hr 17 min \\ \hline
RPIDirectory          & 7 hr 29 min \\ \hline
ShuttleTracking       & 2 hr 2 min \\ \hline
YACS                  & 3 hr 32 min \\ \hline
\end{tabular}
\label{projectdurations}
\end{table}

This grand mean may provide an idea on the scale of the programming tasks one should be assigned to do, as very likely the software project one can achieve could depend on the attention span one can put into the project. Analyzing the average session of successful projects may provide recommended coding time for projects of different scales.

\section{Conclusions}
Sentences and lines of code are rarely grouped together outside of a pseudo code lecture or alike, but the wealth of information they and the meta-data surrounding them provide show that despite their differences in appearance and structure there are many similarities. We have presented the methodology for temporal analysis of literary and programming prose using existing easily-accessible tools. For temporal keywords, we depend on target-oriented searches on the frequency count of related n-grams, assuming frequency count in all the pages as the popularity measure. For commit time, we count the tallies according to the commit time.

As for further work, first we assume occurrence is equivalent popularity. We have not accounted for duplicates that are unrelated to popularity. For more complex analysis, we probably need further programming. We may make use of the APIs offered by Google, for instance, to expand.

In literary works, we are interested in occurrences of temporal keywords across multiple works. Changes across editions of the same work are generally not considered due to the rather limited number of samples with multiple editions. In contrast, temporal keywords seldom occur in source code which makes analysis of their occurrences uninteresting. Rather, we are provided more temporal information such as commit timestamps which make it feasible and meaningful to look into the commit patterns and the changes across different commits. Meanwhile, we are aware of something in-between those two extremes, namely blogs and wiki web pages. They contain not only text but also rich revision information as code repository.


%
\bibliographystyle{abbrv}
\bibliography{paper}  
%
%
\balancecolumns

\end{document}